\documentclass[prb,twocolumn,superscriptaddress,showpacs,preprintnumbers,amsmath,amssymb,floats]
{revtex4}

\usepackage{txfonts}
\usepackage{amssymb}
\usepackage{graphicx}
\usepackage[dvips]{color}

\begin{document}

\title{Significant contribution of As $4p$ orbitals to the low-lying electronic structure of 112-type iron-based superconductor Ca$_{0.9}$La$_{0.1}$FeAs$_2$}

\author{M. Y. Li}
\author{Z. T. Liu}
\affiliation{State Key Laboratory of Functional Materials for Informatics, Shanghai Institute of Microsystem and Information Technology (SIMIT), Chinese Academy of Sciences, Shanghai 200050, China}

\author{W. Zhou}
\affiliation{Department of Physics and Key Laboratory of MEMS of the Ministry of Education, Southeast University, Nanjing 211189, China}

\author{H. F. Yang}
\author{D. W. Shen}\email{dwshen@mail.sim.ac.cn}
\author{W. Li}
\affiliation{State Key Laboratory of Functional Materials for Informatics, Shanghai Institute of Microsystem and Information Technology (SIMIT), Chinese Academy of Sciences, Shanghai 200050, China}

\author{J. Jiang}
\author{X. H. Niu}
\author{B. P. Xie}
\affiliation{State Key Laboratory of Surface Physics, Department of Physics,
and Advanced Materials Laboratory, Fudan University, Shanghai 200433, China}

\author{Y. Sun}
\affiliation{Department of Physics and Key Laboratory of MEMS of the Ministry of Education,
Southeast University, Nanjing 211189, China}

\author{C. C. Fan}
\affiliation{State Key Laboratory of Functional Materials for Informatics, Shanghai Institute of Microsystem and Information Technology (SIMIT), Chinese Academy of Sciences, Shanghai 200050, China}
\author{Q. Yao}
\affiliation{State Key Laboratory of Functional Materials for Informatics, Shanghai Institute of Microsystem and Information Technology (SIMIT), Chinese Academy of Sciences, Shanghai 200050, China}
\affiliation{State Key Laboratory of Surface Physics, Department of Physics,
and Advanced Materials Laboratory, Fudan University, Shanghai 200433, China}
\author{J. S. Liu}
\affiliation{State Key Laboratory of Functional Materials for Informatics, Shanghai Institute of Microsystem and Information Technology (SIMIT), Chinese Academy of Sciences, Shanghai 200050, China}

\author{Z. X. Shi}\email{zxshi@seu.edu.cn}
\affiliation{Department of Physics and Key Laboratory of MEMS of the Ministry of Education,
Southeast University, Nanjing 211189, China}

\author{X. M. Xie}
\affiliation{State Key Laboratory of Functional Materials for Informatics, Shanghai Institute of Microsystem and Information Technology (SIMIT), Chinese Academy of Sciences, Shanghai 200050, China}

\date{\today}

\begin{abstract}
We report a systematic polarization-dependent angle-resolved photoemission spectroscopy study of the three-dimensional electronic structure of the recently discovered 112-type iron-based superconductor Ca$_{1-x}$La$_{x}$FeAs$_2$ ($x$~=~0.1). Besides the commonly reported three hole-like and two electron-like bands in iron-based superconductors, we resolve one additional hole-like band around the zone center and one more fast-dispersing band near the $X$ point in the vicinity of Fermi level. By tuning the polarization and the energy of incident photons, we are able to identify the specific orbital characters and the $k_z$ dependence of all bands. Combining with band calculations, we find As $4p_{z}$ and $4p_{x}~(4p_{y})$ orbitals contribute significantly to the additional three-dimensional hole-like band and the narrow band, respectively. Also, there are considerable hybridization between the As $4p_{z}$ and Fe 3$d$ orbitals in the additional hole-like band, which suggests the strong coupling between the unique arsenic zigzag bond layers and the FeAs layers therein. Our findings provide a comprehensive picture of the orbital characters of the low-lying band structure of 112-type iron-based superconductors, which can be a starting point for the further understanding of their unconventional superconductivity.

\end{abstract}

\pacs{74.25.Jb, 74.70.Xa, 71.20.-b, 79.60.-i}

\maketitle

\section{INTRODUCTION}

In iron-based superconductors, all the five Fe $3d$ bands are involved in the low-lying electronic structure, which is in sharp contrast to the case of cuprate high temperature supercondutors~\cite{Multiband1,Mulitband2,Multiband3}. Theoretically, these additional orbital degrees of freedom have been predicted to result in a strong anisotropy of superconducting gaps for Fermi surface~\cite{gap theory}. This prediction was soon confirmed by various experiments~\cite{gapanisotropy1,gapanisotropy2,gapanisotropy3}. Consequently, the multi-orbital nature is believed to be closely related to the unconventional superconductivity in iron-based superconductors, and it is important to identify experimentally the orbital characters of multiple bands in the vicinity of Fermi energy ($E_F$).

\begin{figure}[!hb]
\includegraphics[width=8.5cm]{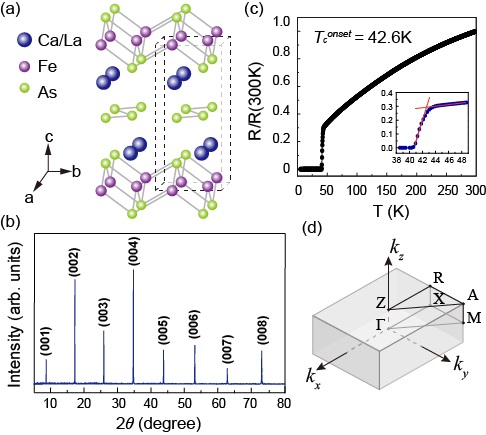}\\
\caption{(Color online). (a) The crystal structure of Ca$_{1-x}$La$_{x}$FeAs$_2$. (b) The X-ray diffraction of Ca$_{0.9}$La$_{0.1}$FeAs$_2$ single crystals. (c) The temperature dependence of normalized resistance in the \emph{ab}-plane of Ca$_{0.9}$La$_{0.1}$FeAs$_2$ single crystal. (d) The schematic drawing of the three-dimensional reduced Brillouin zone of Ca$_{1-x}$La$_x$FeAs$_2$.}
\label{character}
\end{figure}

\begin{figure*}[t]
\includegraphics[width=17.5cm]{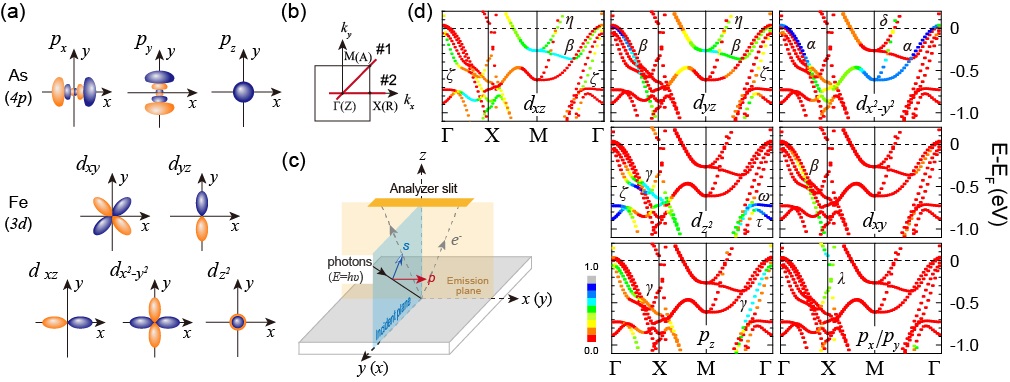}\\
\caption{(Color online). (a) The spatial symmetries of Fe 3$d$ and As 4$p$ orbitals. (b) The two-dimensional projection of the Brillouin zone and the high-symmetry directions. (c) The experimental setup for our polarization-dependent ARPES. The analyzer slit is vertical to the incident plane. The emission plane is defined by the analyzer slit and the sample surface normal. Note that the $s$ polarized light contains a polarization component parallel to \emph{z} axis. (d) The calculated orbital-projected band structure of Ca$_{0.9}$La$_{0.1}$FeAs$_2$. The weight of individual Fe 3$d$ and As $4p$ orbital for different bands is presented by the gradually varied colors.}\label{theory}
\end{figure*}

Recently, a novel family of 112-type iron-based superconductors (Ca, R)FeAs$_2$ (where R = La or Pr) have been reported\cite{112CL,112-CL,112CL1}. Different from other iron pnictides, these new superconductors consist of alternately stacked FeAs and arsenic zigzag bond layers. In addition, by substituting of La for Ca, the superconducting transition temperature ($T_c$) of (Ca, La)FeAs$_2$ can reach as high as 45 K~\cite{112CL,112-CL,112CL1}. This discovery immediately simulated the interest in the possible exotic electronic structure of this new class of iron-based superconductors. Density-functional theory (DFT) calculations soon suggested that there should be four hole-like and two electron-like bands intersecting the Fermi level around $\Gamma$ and $M$ points, respectively, which are mainly derived from the Fe 3$d$ and As 4$p$ orbitals\cite{112CL}. However, this prediction was not in a good agreement with a recent angle-resolved photoemission spectroscopy (ARPES) measurement~\cite{112ES}. Moreover, a recent theoretical work proposed that the unique arsenic zigzag bond layers in CaFeAs$_2$ could generate anisotropic Dirac cones close to the Brillouin zone boundary, to which As $p$ orbitals would contribute substantially ~\cite{112-3D}. This proposal thus calls for a further identification of the orbital characters of the band structure for this class of multi-orbital superconductors. However, to date there have been no such experimental results reported.

In this article, we report a systematic study of the low-lying electronic structure of a typical 112-type superconductor Ca$_{0.9}$La$_{0.1}$FeAs$_2$ ($T_c$~=~42.6 K) using both the polarization-dependent ARPES and band structure calculation. In the vicinity of $E_F$, besides three hole-like and two electron-like bands commonly discovered in other iron pnictides, we resolve one additional hole-like band around the zone center and another fast-dispersing band near the zone boundary. We have further determined the orbital characters and the $k_z$ dependence of these bands by tuning the polarization and the energy of the incident photons. Surprisingly, different from other iron-based superconductors, CaAs layers play a considerable role to the low energy electronic structure of 112-type superconductors. Combining with band structure calculations, our ARPES data suggest a substantial contribution from the As $4p$ orbitals to the additional three-dimensional hole-like band and the fast-dispersing band. Also, there exists hybridization between the As $4p_{z}$ and Fe 3$d$ orbitals in the additional hole-like band, suggesting the strong coupling between FeAs and CaAs blocking layers. Our findings provide a comprehensive picture of the orbital characters of the electronic structure of 112-type iron-based superconductors, which might be a starting point for the further understanding of their unconventional superconductivity.

\section{EXPERIMENTAL AND CALCULATIONAL PROCEDURES }

\begin{figure}[t]
  \includegraphics[width=8.5cm]{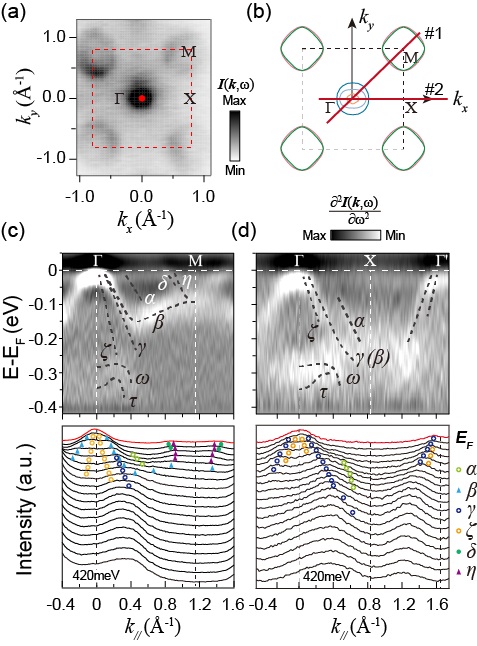}\\
  \caption{(Color online). The electronic structure of Ca$_{0.9}$La$_{0.1}$FeAs$_2$ taken with circularly polarized photons of 104~eV. (a) The photoemission intensity map integrated over [$E_F$ - 10~meV, $E_F$ + 10~meV]. The red dashed lines represent the reduced Brillouin zone. (b) The calculated Fermi surface. (c), (d) Photoemission intensity plots and the corresponding MDCs along $\Gamma$-$M$ and $\Gamma$-$X$, respectively. }\label{FS}
\end{figure}

High quality Ca$_{0.9}$La$_{0.1}$FeAs$_2$ single crystals were synthesized by self-flux method as described elsewhere~\cite{112CL,flux,SEU}. Fig.~\ref{character}(a) illustrates the sketch of the crystal structure of a typical Ca$_{1-x}$La$_x$FeAs$_2$ superconductor, containing alternately stacked FeAs and arsenic zigzag bond layers. This compound keeps the monoclinic structure, different from the common tetragonal or orthotropic crystal structures of most iron-based superconductors~\cite{SEU,112CL}. In Fig.~\ref{character}(b), the X-ray diffraction pattern shows that only the series of (00l) narrow reflection peaks appear, indicating the good crystalline quality of samples. The normalized resistivity in the \emph{ab}-plane gives a $T$$_c$ onset of 42.6 K as illustrated in Fig.~\ref{character}(c)~\cite{SEU}. Its reduced Brillouin zone, corresponding to the unit cell containing two Fe atoms, is shown in Fig.~\ref{character}(d).

The polarization-dependent ARPES has been proven to be a powerful tool to distinguish different orbital characters of bands in iron-based superconductors~\cite{polarize-1, polarize-2, polarize-3, polarize-4, polarize-5}. By specifying the photon polarization, when the analyzer slit is aligned with the high-symmetry directions of samples, the photoemission signal from a certain orbital is expected to appear or disappear, depending on the intrinsic spatial symmetry of this orbital [Fig.~\ref{theory}(a)]. In our DFT calculation, the $x$ and $y$ directions are defined to be along the next nearest Fe-Fe bonds, thus the $\Gamma$-$X$ direction is parallel to $k_x$ ($k_y$) directions in the reciprocal space as indicated by Fig.~\ref{theory}(b). For our setup [Fig.~\ref{theory}(c)], the emission plane, defined by the analyzer slit and the sample surface normal, is vertical to the incident plane. The $p$ and $s$ polarized lights are of electric fields in and out of the emission plane, respectively. Thus, when rotating the sample so as to align the $\Gamma$-$X$ direction with the analyzer slit, $d_{xz}$, $d_{x^2-y^2}$, $d_{z^2}$ are all of even symmetry with respect to the emission plane. Therefore, photoemission signals from these orbitals could be only detected in the \emph{p} geometry. However, for the odd orbitals $d_{xy}$ and $d_{yz}$, their photoemission signals appear only in the $s$ geometry. Likewise, we could deduce the results when the slit is along the $\Gamma$-$M$ direction. Note that $d_{xz}$ and $d_{yz}$ are not symmetric with respect to the emission plane in this case, and thus could be observed in both the $p$ and $s$ geometries. Also, As $p_z$ orbital shares the same spacial symmetry with Fe $d_{z^2}$ with respect to the emission plane. Here, we emphasize that we used the Fe atomic spatial symmetry to infer the the orbital characters of bands. However, this simple picture is accurate enough for the bands around the zone center~\cite{Brouet}. In this way, the possibility to detect different orbitals in the $p$ and $s$ geometries is summarized in Table~\ref{tab:table1}.

Most of our polarization-dependent ARPES measurements were conducted at the beamline I05 of Diamond light source (DLS) with the linearly polarized photons. The data were taken at 7 K in a ultrahigh vacuum better than 8$\times$10$^{-11}$ torr. The circularly polarized data were taken at 15$\sim$20 K in a ultrahigh vacuum of around 3$\times$10$^{-11}$ torr at the SIS Beamline of Swiss Light Source (SLS). Both setups are equipped with VG-Scienta R4000 electron analyzers. The angular resolutions were set to 0.2$^{\circ}$ and 0.3$^{\circ}$ for DLS and SLS, respectively. The overall energy resolution was set to 10$\sim$30~meV depending on the photon energy. Samples were cleaved ${in}$-${situ}$ and exposed the natural cleaved ${ab}$-plane. During the measurements the samples were stable and did not show any sign of degradation.

The electronic band calculations were performed within the density functional formalism as implemented in the VASP code~\cite{VASP}. The plane wave basis method and the Perdew-Burke-Ernzerhof exchange correlation potential were used~\cite{PBE}. The lattice constants employed in this theoretical calculation are \emph{a}~=~3.9471{\AA}, \emph{b}~=~3.8724{\AA} and \emph{c}~=~10.3210{\AA}, which were obtained from the single-crystal X-ray diffraction analysis~\cite{112CL}. Throughout the calculation, a $500$ eV cutoff in the plane wave expansion and a $9\times 9\times 9$ Monkhorst-Pack $\vec{k}$ grid are chosen to ensure the calculation with an accuracy of $10^{-5}$ eV.

\begin{figure*}[t]
\includegraphics[width=17cm]{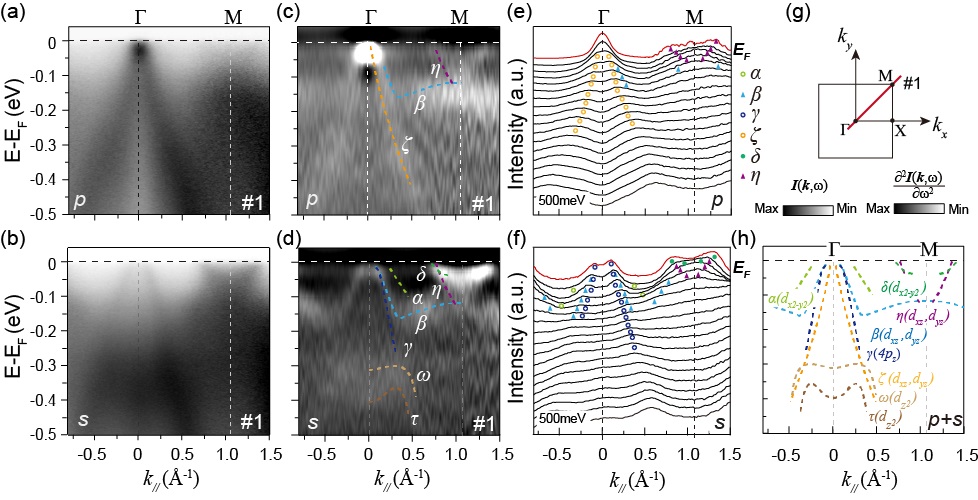}\\
  \caption{(Color online). The polarization-dependent photoemission data along $\Gamma$-$M$ taken with 104 eV photon energy. (a), (b) The photoemission intensity plots along $\Gamma$-$M$ in the $p$ and $s$ geometries, respectively. (c), (d) The second derivative with respect to energy corresponding to (a) and (b), respectively. (e), (f) The MDCs for the data in panel (a) and (b), respectively. (g) The indication of the cut direction in the projected two-dimensional Brillouin zone. (h) The summary of the bands detected along $\Gamma$-$M$ in both the $p$ and $s$ geometries. }\label{cut1}
\end{figure*}

\begin{table}[t]
\caption{\label{tab:table1} The possibility of detecting Fe ${3d}$ orbitals along two high-symmetry directions in the $p$ and $s$ geometries by polarization-dependent ARPES.}
\begin{ruledtabular}
\begin{tabular}{c|cccccc}
High symmery   &  \multicolumn{5}{c}{ Fe 3$d$ orbitals} & \\
directions   &  &   &    &   &  & \\
\hline
\ $\Gamma$-$M$~~( $p$ )  & $d_{xz}$    &   $d_{yz}$   &   $d_{xy}$    &   &  $d_{z^2}$  & \\
\hline
\ $\Gamma$-$M$~~( $s$ )  & $d_{xz}$    &   $d_{yz}$   &       & $d_{x^2-y^2}$   &     & \\
\hline
\ $\Gamma$-$X$~~( $p$ )  & $d_{xz}$    &       &   &  $d_{x^2-y^2}$   &   $d_{z^2}$  & \\
\hline
\ $\Gamma$-$X$~~( $s$ )  &     &   $d_{yz}$   &   $d_{xy}$    &    &    & \\
\hline

\end{tabular}
\end{ruledtabular}
\end{table}

\section{RESULTS AND DISCUSSION}

\subsection{The low-lying electronic structure of Ca$_{0.9}$La$_{0.1}$FeAs$_2$}

According to the calculation [Fig.~\ref{theory}(d)], along the \emph{$\Gamma$}-$M$ direction, both $d_{xz}$ and $d_{yz}$ orbitals make a significant contribution to the innermost hole-like band $\zeta$. Besides, they form the hole-like band $\beta$ and the electron-like band $\eta$, which are degenerate at $M$. Another group of hole- and electron-like bands ($\alpha$ and $\delta$), which both originate from $d_{x^2-y^2}$ orbital, are degenerate at $M$ as well, but at a higher binding energy. As for the $d_{z^2}$ orbital, it mainly contributes to $\omega$ and $\tau$ bands, which are well below the Fermi level. We note As 4${p_z}$ orbital hybridizes with Fe 3$d_{z^2}$, forming another hole-like band $\gamma$ around the $\Gamma$ point. Moreover, As 4${p_x}$ (${p_y}$) orbitals contribute to the $\lambda$ band near $X$ along $M$-$X$.

\begin{figure*}[t]
\includegraphics[width=17cm]{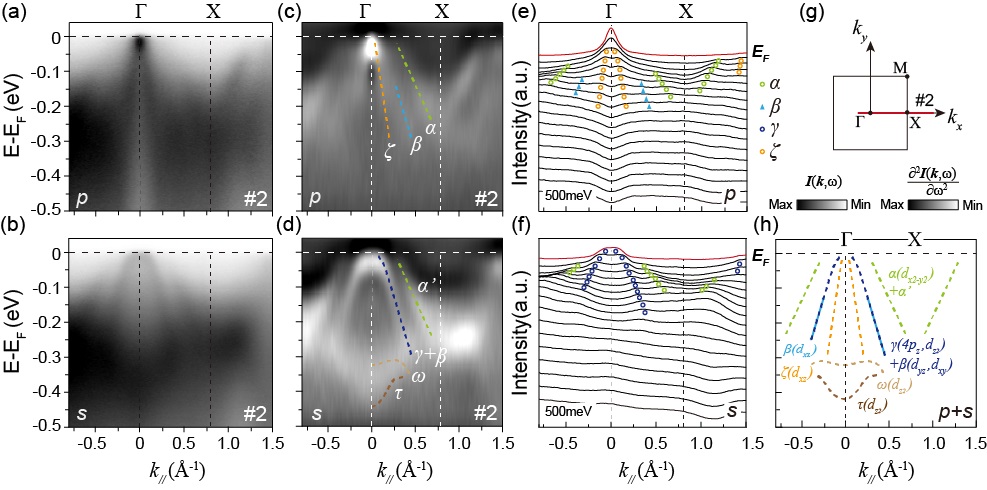}\\
  \caption{(Color online). The polarization-dependent photoemission data along $\Gamma$-$X$ taken with 104 eV photon energy. (a), (b) The photoemission intensity plots along $\Gamma$-$X$ in the $p$ and $s$ geometries, respectively. (c), (d) The second derivative with respect to energy corresponding to (a) and (b), respectively. (e), (f) The MDCs for the data in panel (a) and (b), respectively. (g) The indication of the cut direction in the projected two-dimensional Brillouin zone. (h) The summary of the bands detected along $\Gamma$-$X$ in both the $p$ and $s$ geometries. }\label{cut2}
\end{figure*}

The overall electronic structure of Ca$_{0.9}$La$_{0.1}$FeAs$_2$ is shown in Fig.~\ref{FS}. Fig.~\ref{FS}(a) shows the photoemission intensity map taken with the circularly polarized photons. Qualitatively, the resulting Fermi surface topology agrees well with the band structure calculation [Fig.~\ref{FS}(b)]. However, it is difficult to distinguish these predicted multiple bands directly from the map due to the overwhelming intensity of some broad bands. To better resolve these bands, we present both the second derivative with respect to energy and the corresponding momentum distribution curves (MDCs) of the photoemission intensity taken along $\Gamma$($Z$)-$M$($A$) (cut~\#1) and $\Gamma$($Z$)-$X$($R$) (cut~\#2) in Figs.~\ref{FS}(c) and (d), respectively. For cut~\#1 [Fig.~\ref{FS}(c)], we can infer that the outermost hole-like band around $\Gamma$($Z$) should be $\alpha$, and the adjacent ones are the nearly degenerate $\beta$ and $\gamma$ bands. $\beta$ starts to bend over towards $M$($A$) at around 150~meV below $E_F$, and then degenerates with the inner electron-like band $\eta$. Finally, the outer electron-like band around $M$($A$) and the innermost hole-like band around $\Gamma$($Z$) could be assigned to $\delta$ and $\zeta$, respectively, according to the DFT calculation. In a similar way, all the bands taken along $\Gamma$($Z$)-$X$($R$) can be determined, as illustrated in Fig.~\ref{FS}(d). Note that the $\gamma$ and $\beta$ bands seem to be degenerate with each other along this direction and we cannot resolve them within our experimental resolution. However, taking advantage of their different spacial symmetries, we could distinguish them with linearly polarized photons hereinafter [Figs.~\ref{cut1} and~\ref{cut2}].

Below $E_F$, there are two bands around the zone center, which could be assigned to $\omega$ and $\tau$ as predicted by DFT. Because these two bands are relatively flat, it is difficult to track their dispersions from MDCs. However, by fitting the peak positions of the corresponding energy distribution curves (not shown), we could still obtain their dispersions quantitatively, as shown in Figs.~\ref{FS}(c) and (d). In general, our experimental results show remarkable consistence with the DFT calculation, which from one side confirms that the electronic structure measured by ARPES could well represent the bulk information of Ca$_{0.9}$La$_{0.1}$FeAs$_2$.

\subsection{Orbital characters of bands near $E_F$}
In order to determine the orbital characters of all the low-lying bands in Ca$_{0.9}$La$_{0.1}$FeAs$_2$, we next performed the polarization-dependent ARPES measurements along two high-symmetry directions. Fig.~\ref{cut1} shows the results taken along cut~\#1~($\Gamma$-$M$) in both geometries. In the vicinity of $E_F$, there are two bands (the light blue and purple dashed lines) which show up simultaneously in both the $p$ and $s$ geometries, as illustrated by both the second derivative data and the corresponding MDCs [Figs.~\ref{cut1}(c-f)]. According to Table~\ref{tab:table1}, these two bands could only originate from $d_{xz}$ and $d_{yz}$ orbitals, which have no definite symmetry with respect to $\Gamma$-$M$. Consequently, we could assign them as the hole-like band $\beta$ and the electron-like band $\eta$, respectively, which have been predicted to originate from a mixture of $d_{xz}$ and $d_{yz}$ orbitals [Fig.~\ref{theory}(d)].

Besides $\beta$, there appears one more hole-like band with faster Fermi velocity around $\Gamma$ in the $p$ and $s$ geometries, respectively, highlighted by the orange and dark blue dashed lines in Figs.~\ref{cut1}(c) and (d). By extracting the peak positions of MDCs, we could obtain the band dispersion in each geometry [Figs.~\ref{cut1}(e) and (f)]. The differences in both the Fermi velocity and Fermi crossing indicate that they are two distinct bands. Referring to the DFT calculation in Fig.~\ref{theory}(d), they should be either $\zeta$ or $\gamma$ band. For the $p$ geometry, the hole-like band (the orange one) crosses the Fermi level barely and disperses a little faster, and thus we could assign it to be the $\zeta$ band, which was predicted to be of both $d_{xz}$ and $d_{yz}$ characters. Consequently, another hole-like band appearing in the $s$ geometry (the dark blue one) could be assigned as $\gamma$, mainly from the As $4p_z$ orbital according to the DFT calculation. Here we note that the $\zeta$ band only shows up in the $p$ geometry although the $d_{xz}$ and $d_{yz}$ orbitals have no specified symmetry with respect to $\Gamma$-$M$. Such strong polarization dependence is mostly likely due to the hybridization of the $d_{xz}$ and $d_{yz}$ orbitals, which might form states of pure even or odd spatial symmetry with respect to $\Gamma$-$M$, similar to the case of BaFe$_{1.85}$Co$_{0.15}$As$_2$ \cite{polarize-3}.

\begin{figure*}[t]

\includegraphics[width=15cm]{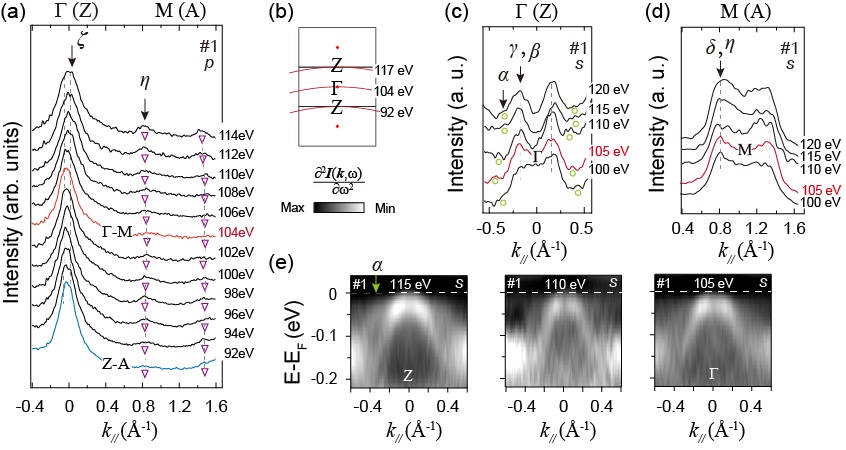}\\
  \caption{(Color online). Photon-energy dependence of bands along $\Gamma$($Z$)-$M$($A$). (a) The photon-energy dependence of the MDCs near $E_F$ in the $p$ geometry. (b) The correspondence between the typical photons and the high-symmetry points along $k_z$ direction in the three-dimensional Brillouin zone. (c) The photon-energy dependence of the MDCs near $E_F$ around $\Gamma$($Z$) in the $s$ geometry. (d) The photon-energy dependence of the MDCs at $E_F$ around $M$($A$) in the $s$ geometry. (e) The second derivatives with respect to energy taken along $\Gamma$($Z$)-$M$($A$) in the $s$ geometry with 105 eV, 110 eV and 115 eV photons, respectively. }\label{kz1}
\end{figure*}

In the $s$ geometry, we could identify two more bands near $E_F$, one hole-like band around $\Gamma$ and another electron-like band around $M$, both of which appear only in the $s$ geometry. Here, the hole-like feature is weak and relatively difficult to resolve in the 104~eV photon data shown in Fig.~\ref{cut1}. While, it could be better distinguished in other $k_z$ planes and would be discussed later in Fig.~\ref{kz1}(e). As suggested by Table~\ref{tab:table1}, the only possible orbital which might contribute to these bands is $d$$_{x^2-y^2}$. Moreover, the calculation has shown that $d$$_{x^2-y^2}$ indeed contributes weight to the hole-like band $\alpha$ and the electron-like band $\delta$ along $\Gamma$-$M$. We could thus assign them as $\alpha$ and $\delta$, respectively. Around $\Gamma$, there are some spectral feature at binding energy below 300~meV. After referring to the DFT calculation, we could attribute the spectral weight to the bands $\omega$ and $\tau$, which both mainly come from the $d_{z^2}$ orbital. We note that these two bands are visible in the $s$ geometry but suppressed in the $p$ geometry. This might be due to the significant enhancement of the photoemission signal for the $d_{z^2}$ orbital induced by the component of the $s$ polarized light along the $z$ axis~\cite{polarize-4}. Actually, the similar enhancement of $d_{z^2}$ orbital has been reported in a recent polarization-dependent ARPES work on (Tl,~Rb)$_y$Fe$_{2-x}$As$_2$, in which the same experimental setup as ours was used~\cite{polarize-5}. In the same way, we can attribute the similar polarization selectivity of the $\gamma$ band to its considerable As $4p_z$ orbital character. We can summarize all the identified bands together with their orbital characters along cut~\#1 in Fig.~\ref{cut1}(h).

The photoemission data taken along cut~\#2 ($\Gamma$-$X$) are shown in Fig.~\ref{cut2}. Along this direction, the $p$ geometry data [Figs.~\ref{cut2}(a), (c) and (e)] show that there are three hole-like bands around $\Gamma$ near $E_F$. As suggested by Table~\ref{tab:table1}, the candidates of the orbitals should be $d_{xz}$ and $d_{x^2-y^2}$. Note that the $d_{z^2}$ is almost irrelevant to the low-lying band structure as predicted in Fig.~\ref{theory}(d). Thus, the outermost hole-like band around $\Gamma$ could be assigned as $\alpha$, which has the $d_{x^2-y^2}$ orbital component. As for the innermost hole-like band, in view of its barely Fermi crossing and fastest dispersion, we could attribute this band to $\zeta$, which is of substantial $d_{xz}$ orbital character. These assignments above are well consistent with the polarization-dependent ARPES results along cut~\#1. In between $\zeta$ and $\alpha$, the third hole-like band could be either $\beta$ or $\gamma$. Since $\beta$ has been determined to originate from both $d_{xz}$ and $d_{yz}$ orbitals in cut~\#1 above, this band can be assigned as $\beta$.

Next we could assign the inner feature appearing in the $s$ geometry as $\beta$ since it shares the same dispersion with the middle band in the $p$ geometry. However, considering the enhancement of $d_{z^2}$ (${p_z}$) orbital for the $s$ geometry, we could not exclude the contribution to the inner feature from $\gamma$ , which is expected to be the mixture of Fe $3d_{z^2}$ and As ${4p_z}$ according to the DFT calculation. So we can deduce that $\beta$ and $\gamma$ are probably degenerate along this direction. Due to the same enhancement of $d_{z^2}$ orbital, we detected the bands $\omega$ and $\tau$ as well, in good agreement with the case of cut~\#1. Moreover, we can identify another hole-like band around $\Gamma$, which shares the same dispersion with $\alpha$ in the $p$ geometry. Naturally, we would like to attribute this feature to $\alpha$ band as well. However, according to the DFT calculation, the $\alpha$ band is of pure $d_{x^2-y^2}$ orbital character, and it is not expected to show up in the $s$ geometry along $\Gamma$-$X$. One possibility is that there are still some other orbital characters apart from $d_{x^2-y^2}$, which might bring the photoemission spectral weight in the $s$ geometry. We will denote this feature as $\alpha'$ hereafter. To analyze the detailed orbital characters of this band, we next performed the photon-energy dependent ARPES measurements [Figs.~\ref{kz1} and~\ref{kz2}]. As a summary for this part, we append all the identified bands and their orbital characters along cut~\#2 in Fig.~\ref{cut2}(h).

\subsection{$k_z$ dependence of the band structure}

Fig.~\ref{kz1} shows the $k_z$ dependence along $\Gamma$($Z$)-$M$($A$) direction for different photon energies in the $p$ and $s$ geometries. Here, the inner potential of 15~eV was taken to estimate the $k_z$'s for different photon energies. Fig.~\ref{kz1}(b) sketches the correspondence between the typical photons we applied and the high-symmetry points along $k_z$ direction in the three-dimensional Brillouin zone of Ca$_{0.9}$La$_{0.1}$FeAs$_2$. In Fig.~\ref{kz1}(a) stack the MDCs of the photoemission spectra near $E_F$ in the $p$ geometry. The peak positions of both bands $\zeta$ and $\eta$ show negligible $k_z$ dependencies, indicating their rather two-dimensional nature. However, the story is a little different in the $s$ geometry along the same cut [Figs.~\ref{kz1}(c) and (d)]. While the hole-like bands $\beta$ and $\gamma$, electron-like bands $\delta$ and $\eta$ exhibit two-dimensional characters, the Fermi crossing of $\alpha$ shows pronounced $k_z$ dependence, suggestive of considerable three-dimensionality of its band dispersion. Note that there are a little changes of the relative intensity of bands with different photon energies due to the matrix element effects. For example, the faint feature of $\alpha$ band was detected around $\Gamma$, nevertheless the feature looked more evident around $Z$ as shown in Fig.~\ref{kz1}(e).

\begin{figure}[t]
\includegraphics[width=8.5cm]{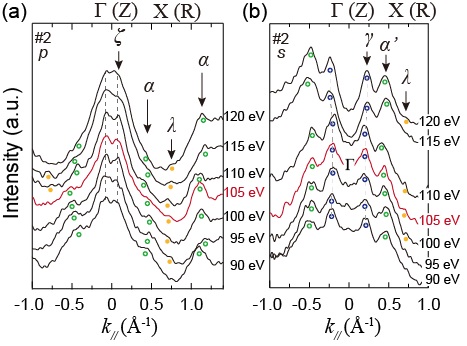}\\
  \caption{(Color online). Photon-energy dependence of bands along $\Gamma$($Z$)-$X$($R$). (a) The photon-energy dependence of the MDCs near $E$$_F$ in the $p$ geometry. (b) The photon-energy dependence of the MDCs near $E$$_F$ in the $s$ geometry. }\label{kz2}
\end{figure}

Also, we performed the photon-energy dependent ARPES measurements along $\Gamma$($Z$)-$X$($R$) direction. As shown by the MDCs of the photoemission spectra near $E_F$ in both geometries [Figs.~\ref{kz2}(a) and (b)], we could conclude that nearly all the bands in the vicinity of $E_F$ show two-dimensional nature except $\alpha$ shows considerable three-dimensional behavior. In view of the three-dimensional character, $\alpha$ should not merely originate from the pure $d_{x^2-y^2}$ orbital as predicted by the current DFT calculation. The $d_{x^2-y^2}$ must mix with some other more three-dimensional orbital components to form the $\alpha$ and $\alpha'$ bands. A recent theoretical calculation on CaFeAs$_2$ proposed that the Ca $d$ and As $4p_z$ orbitals would contribute to the outermost hole pocket around the zone center besides Fe $3d$ orbitals, which would induce the strong three-dimensionality of the dispersion~\cite{112-3D}. Therefore, a complex mixture of Fe 3$d_{x^2-y^2}$, As 4$p_z$ and Ca $d$ orbitals might be a more reasonable description of the orbital character for the $\alpha$ band, and the $\alpha'$ band in the $s$ geometry might originate from As 4$p_z$ and Ca $d$ orbitals.

\subsection{The additional fast-dispersing band discovered near the zone boundary}

\begin{figure}[t]
\includegraphics[width=9cm]{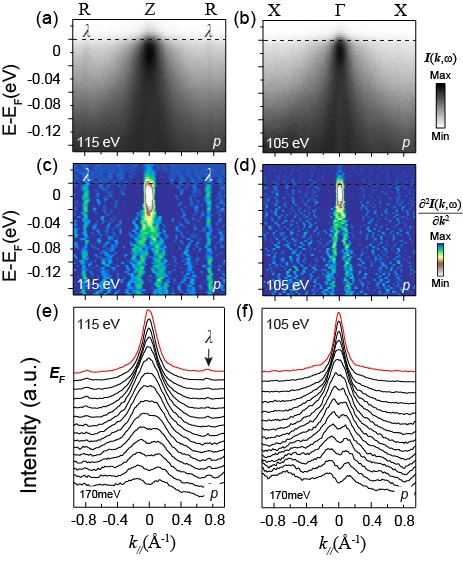}
\caption{(Color online). The photoemission evidence of the additional band $\lambda$ near the zone boundary. (a), (b) The photoemission intensity plots taken along $\Gamma$($Z$)-$X$($R$) in the $p$ geometry with 105 eV and 115 eV photons, respectively. (c), (d) The second derivatives relative to the momentum for photoemission intensity plots in (a) and (b), respectively. (e), (f) The corresponding MDCs for the data in (a) and (b), respectively. }\label{discuss}
\end{figure}

By tuning the photon energies we could observe an additional fast-dispersing band ($\lambda$) near the zone boundary in the $p$ geometry [Fig.~\ref{discuss}(a)]. This feature can be well identified in the ${ZRA}$ plane, but its intensity becomes rather faint around the $X$ point [Fig.~\ref{discuss}(b)], which is probably due to the differences in photoemission matrix elements. This finding could be further confirmed by both the second derivatives relative to the momentum [Fig.~\ref{discuss}(c) and (d)] and the corresponding MDCs [Fig.~\ref{discuss}(e) and (f)]. Both our calculation [Fig.~\ref{theory}(d)] and the recent DFT work by X. X. Wu ${et~al.}$~\cite{112-3D} have predicted that the arsenic zigzag bond layers would generate an anisotropic Dirac cone close to the $X$ point and it should be mainly contributed by As 4$p_x$ and 4$p_y$ orbitals. These theoretical predictions are in qualitatively agreement with our finding except that the Dirac cone like band dispersion could not be confirmed. Moreover, the predicted As $4p_x$ and $4p_y$ orbital character of this band is consistent with the two-dimensional behavior of $\lambda$ shown in the photon-energy dependent measurement [Fig.~\ref{kz2}(a)]. This finding shows that the arsenic zigzag bond layers are metallic in the 112-type iron-based superconductors, which as well induce the unique electronic properties differing from the other iron-based superconductors.

\section{Conclusion}

In summary, we report a comprehensive study of the orbital characters of the low-lying electronic structure of a typical 112-type superconductor Ca$_{0.9}$La$_{0.1}$FeAs$_2$ by both the polarization-dependent ARPES and the band structure calculation. We have determined the orbital characters and the $k_z$ dependence of the band structure by tuning the polarization and the energy of the incident photons. Surprisingly, because of the extra arsenic zigzag bond layers, CaAs layers are strongly coupled with FeAs layers, resulting in two unique bands contributed by As $p$ orbitals significantly. Our findings lay out a comprehensive picture of the orbital characters of the electronic structure of 112-type iron-based superconductors, which might provide a starting point for the further understanding of their unconventional superconductivity.

\section{ACKNOWLEDGMENTS}
We gratefully acknowledge the experimental support by Dr. P. Dudin at DLS and Dr. M. Shi at SLS, and the helpful discussion with Prof. Donglai Feng and Dr. Xiaoping Wang. This work was supported by National Basic Research Program of China (973 Program) under the grant Nos. 2011CBA00106 and 2012CB927400, and the National Science Foundation of China under Grant Nos. 11104304, 11274332, 11227902 and U1432135. M. Y. Li and D. W. Shen are also supported by the "Strategic Priority Research Program (B)" of the Chinese Academy of Sciences (Grant No. XDB04040300).

\end{document}